# Formation and Trapping of $CO_2$ from Cryogenic Irradiation of Carbonate

Ashma Pandya , Swaroop Chandra , and Michael E. Brown
Division of Geological and Planetary Sciences, California Institute of Technology, Pasadena, CA 91125, USA; apandya@caltech.edu

## Abstract

The detection of $CO_2$ on the Jovian satellite Europa by Galileo/NIMS and recent mapping of the leading side by JWST has revealed that it is most concentrated in geologically young terrains, and its $\nu_3$ asymmetric stretch appears as a spectral doublet centered at 4.25 and 4.27 $\mu$m. Since crystalline $CO_2$ is unstable at Europan surface conditions, this observation implies an active source and a trapping medium, which may be separate. To this end, several hypotheses have been proposed, but no laboratory work has successfully reproduced the spectral features of $CO_2$ on Europa so far. Radiolyzed carbonates have also been discussed as plausible precursors and host materials for $CO_2$, though their role has not been experimentally validated in a Europa-like environment. Here, we report the first laboratory experiments investigating $CO_2$ production from carbonate salts exposed to 10 keV electron irradiation at 50, 100, and 120 K in ultrahigh vacuum. Using diffuse reflectance FTIR spectroscopy, we observe the emergence, growth, and saturation of an absorption doublet centered near 4.25 and 4.27 $\mu$m, consistent with the $CO_2$ $\nu_3$ band. Postirradiation thermal desorption studies using residual gas analysis reveal that the radiolytically formed $CO_2$ is stable at temperatures beyond Europa's surface. This work provides the first experimental evidence that low-energy electron irradiation of carbonates in cryogenic, vacuum conditions can produce and retain $CO_2$, and suggests that carbonates can serve as endogenous reservoirs of $CO_2$ on irradiated icy bodies in the outer solar system.

*Unified Astronomy Thesaurus concepts:* Laboratory astrophysics (2004); Europa (2189); Surface composition (2115); Galilean satellites (627); Planetary surfaces (2113); Carbon dioxide (196)

## 1. Introduction

Icy bodies in the outer solar system have been studied extensively by missions such as Galileo, Cassini, and, more recently, JWST, revealing a diverse inventory of volatile species. $CO_2$ is one of the more prominent volatiles, offering clues about surface composition, radiation chemistry, and potential subsurface processes. Among the $CO_2$-bearing objects, the Jovian moon Europa is particularly interesting, since it not only hosts a subsurface ocean (J. D. Anderson et al. 1998; M. G. Kivelson et al. 2000) that may support habitable conditions, but also displays complex geological activity that appears to couple the surface to the ocean below (R. W. Carlson et al. 2009).

$CO_2$ on Europa was first detected by Galileo/NIMS via its strong $\nu_3$ asymmetric stretch near 4.26 $\mu$m (T. B. McCord et al. 1998). More recently, JWST/NIRSpec mapping has resolved this feature into a spectral doublet centered at 4.249 and 4.268 $\mu$m (S. K. Trumbo & M. E. Brown 2023; G. L. Villanueva et al. 2023). However, this detection poses a thermodynamic paradox: crystalline $CO_2$ sublimes at 80 K in vacuum (C. Ahrens et al. 2022), while Europa's surface reaches temperatures up to 120 K. Its continued presence implies that $CO_2$ is either being actively produced or stabilized by interactions with host materials (T. B. McCord et al. 1998). The observed band centers also deviate from that of pure $CO_2$ ice at 4.268 and 2.698 $\mu$m (A. Oancea et al. 2012), suggesting molecular geometry perturbations that alter vibrational modes and shift absorption features. Further, while $CO_2$ has also been identified on the other Galilean satellites (T. B. McCord et al. 1998; C. A. Hibbitts et al. 2000, 2003; D. Bockelée-Morvan et al. 2024; R. J. Cartwright et al. 2024), the Saturnian satellites (B. J. Buratti et al. 2005; R. N. Clark et al. 2005, 2008; R. H. Brown et al. 2006; D. P. Cruikshank et al. 2010; M. E. Brown et al. 2025), and distant icy objects (L. A. Lebofsky 1975; M. E. Brown & W. C. Fraser 2023; I. Wong et al. 2024; M. N. De Prá et al. 2025), Europa's $CO_2$ is unique in its strong spatial association with geologically disrupted terrain, such as Tara Regio (S. K. Trumbo & M. E. Brown 2023).

This spectral behavior and spatial pattern have spurred several hypotheses for the source and retention mechanism for $CO_2$. Endogenous production through radiolysis is considered a strong possibility, considering Europa's exposure to intense ion and electron bombardment within Jupiter's magnetosphere. Laboratory studies involving irradiation have focused on $CO_2$ generation from hydrocarbons (V. Mennella et al. 2004; O. Gomis & G. Strazzulla 2005; R. W. Carlson et al. 2009; K. P. Hand & R. W. Carlson 2012; G. L. Villanueva et al. 2023), though there is no evidence of organics on Europa (S. K. Trumbo & M. E. Brown 2023; G. L. Villanueva et al. 2023). Other mechanisms have also been discussed, including delivery by impactors (R. W. Carlson et al. 2009), production from a subsurface ocean, and thermal or radiolytic release from carbon-rich minerals (T. B. McCord et al. 1998; C. A. Hibbitts & J. Szanyi 2007; D. Fulvio et al. 2012; U. Raut et al. 2012; S. K. Trumbo & M. E. Brown 2023). Laboratory efforts to characterize the retention of $CO_2$ have hypothesized trapping in water ice (S. A. Sandford & L. J. Allamandola 1990; O. Gálvez et al. 2008; A. Oancea et al. 2012; D. Bockelée-Morvan et al. 2024; L. Schiltz et al. 2024; B. D. Mamo et al. 2025), physisorption on clay minerals (C. A. Hibbitts & J. Szanyi 2007), and entrapment within nonice materials and mixtures (T. B. McCord et al. 1998; O. Gálvez et al. 2008). Despite extensive modeling and experimentation, no single scenario has successfully reproduced both the band shape and central wavelengths of Europa's observed $CO_2$.

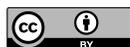

One candidate mechanism is the radiolytic production of $CO_2$ from carbonate salts (D. Bockelée-Morvan et al. 2024; R. J. Cartwright et al. 2024). This pathway is geochemically plausible since carbonates are expected to form in alkaline aqueous environments, as predicted by models of Europa's early ocean (M. Melwani Daswani et al. 2021). Additionally, the spectrum of Tara Regio has a weak, broad 3.9 $\mu$m absorption feature that resembles carbonate-rich deposits identified in the Occator crater on Ceres (F. G. Carrozzo et al. 2018; S. K. Trumbo & M. E. Brown 2023). Although carbonates have not been definitively detected on Europa's surface, this subtle 3.9 $\mu$m feature makes it difficult to eliminate their presence, especially as remnants of past ocean–surface exchange. Carbonates are well known to thermally decompose to release $CO_2$, but whether they undergo radiolytic decomposition under Europa-like conditions has not been experimentally tested.

In this study, we sought to test the hypothesis that radiolytic decomposition of carbonates could represent a viable, endogenous $CO_2$ production pathway on icy, irradiated bodies like Europa, using calcium carbonate ($CaCO_3$) as a model system. The radiation response of $CaCO_3$ has previously been investigated under high-energy irradiation using electron paramagnetic resonance (EPR) spectroscopy (C. Jacobs et al. 1989; R. Debuyst et al. 1993; A. Negrón-Mendoza et al. 2015, 2016). Recent work by I. G. Zhevtun et al. 2024 found that the IR spectrum of $CaCO_3$ remains almost unchanged after 30 keV electron irradiation at 300 K. Here, we irradiated $CaCO_3$ samples at low temperatures using low-energy (10 keV) electrons under ultrahigh vacuum (UHV) and monitored $CO_2$ formation and postirradiation temperature desorption using FTIR and residual gas analysis (RGA). Our work shows that $CaCO_3$ irradiated at cryogenic temperatures releases $CO_2$ with temperature-dependent behavior that may provide insight into Europa's surface $CO_2$ distribution.

## 2. Experimental Setup

We performed a series of electron irradiation experiments to investigate radiation-induced spectral changes in calcium carbonate (ACS grade, CAS No. 12471-34-1) at low temperatures, specifically in the 4–4.5 $\mu$m region relevant to $CO_2$ absorption. The experimental setup is meant to simulate Europa's surface, where substrates experience irradiation in cryogenic, UHV conditions. $CaCO_3$ was chosen since it does not have any intrinsic spectral features in the 4.2–4.3 $\mu$m region, enabling unambiguous detection of newly formed $CO_2$. Although calcium has not been detected on Europa's surface, the presence of $Ca^+$ has been inferred from sputtered ion detections (M. Volwerk et al. 2001). This species could have originated on the seafloor (M. Melwani Daswani et al. 2021) and could be delivered to the surface in a manner analogous to sodium and potassium (O. Ozgurel et al. 2018).

In each experiment, fine calcium carbonate powder ($<$ 100 $\mu$m grains) was pressed into a 0.127 mm thick indium foil placed inside a copper sample cup with an inner diameter of 14 mm and depth of 1 mm. The sample cup was tapped to dislodge unembedded grains, leaving behind a thin layer of salt implanted in the foil. The sample assembly was mounted onto a cold finger within a Kimball Physics UHV chamber. The chamber was evacuated to $\sim 10^{-8}$ Torr using an Agilent ID3 backing pump and a TwisTorr 84 molecular turbo pump. The sample was maintained at 300 K for about 5 hr to remove extant $CO_2$ present on the sample and cold finger, cooled to the experimental temperature using an ARS-4HW water-cooled helium compressor and held for another 5 hr. The sample was then irradiated for 6 hr with 10 keV electrons from a Kimball Physics EGG 301 electron gun. The electron beam current was measured using a Kimball Physics Faraday cup mounted on a linear actuator, which could be positioned in the beam path. After irradiation, the sample was warmed to 150 K at 0.2 K minute$^{-1}$ using a Lake Shore Model 335 temperature controller to identify the sublimation temperature of any $CO_2$ formed.

A Nicolet iS50 FTIR spectrometer was used to collect infrared spectra from 1.3 to 8 $\mu$m with a resolution of 0.241 cm$^{-1}$. The spectra were collected right before irradiation, then continuously throughout the 6 hr of irradiation, then immediately after turning off the electron beam. Each spectrum was an average of 630 scans, taken over 10 minutes. Infragold from Labsphere was used as the reference standard. Diffuse reflectance was measured 90° from the specular direction and at a 45° incidence angle. The optical path outside the vacuum chamber was continuously flushed with nitrogen gas to minimize telluric interference. To monitor gaseous release during the irradiation and warm-up phases, we used a Kurt J. Lesker RGA consisting of a quadrupole-based mass spectrometer. Mass spectra over the 1–50 amu range were taken continuously with a dwell time of 256 ms and resolution of 0.1 amu, from the time the sample reached the irradiation temperature until the postirradiation warming was complete.

Since crystalline $CO_2$ sublimes at 80 K in UHV, we conducted initial experiments at 50 K to minimize $CO_2$ loss by outgassing during irradiation. The initial current density was 4244 nA cm$^{-2}$, corresponding to an electron flux of 2.6 × 10$^{13}$ electrons cm$^{-2}$ s$^{-1}$. Since 10 keV electrons penetrate 1 $\mu$m (H. S. Wong & N. R. Buenfeld 2006) in $CaCO_3$, this flux corresponds to a dose rate of 8.2 × 10$^5$ eV (16 amu yr)$^{-1}$. Compared to Europa's leading hemisphere, which has a dose rate of 10 eV (16 amu yr)$^{-1}$ (T. A. Nordheim et al. 2018), this is about 82,000 times higher. Over 6 hr of irradiation, the energy deposited on the sample (i.e., the absorbed dose) is equivalent to approximately 56 yr on Europa. We then conducted a second experiment using one-third of the initial current density to test whether the production and retention of radiolytic $CO_2$ depend only on the total absorbed dose or also the rate at which energy is deposited. Additional experiments at 100 and 120 K at the high-flux setting were also performed to explore this chemistry in the context of Europa's surface temperatures.

For each of the four experiments, at least three trials were performed. In the figures below, the spectra with minimum noise and background $CO_2$ effects, and flat, stable baselines are shown, along with RGA traces that spanned the entire duration of the experiment and had the highest temperature resolution. Control experiments included irradiation of an empty sample cup lined with indium foil, and thermally ramping an unirradiated $CaCO_3$ sample, initially held at 50 K, to 150 K in UHV. As shown below, these tests confirmed that the spectral results were not influenced by indium or the potential thermal decomposition of the salt.

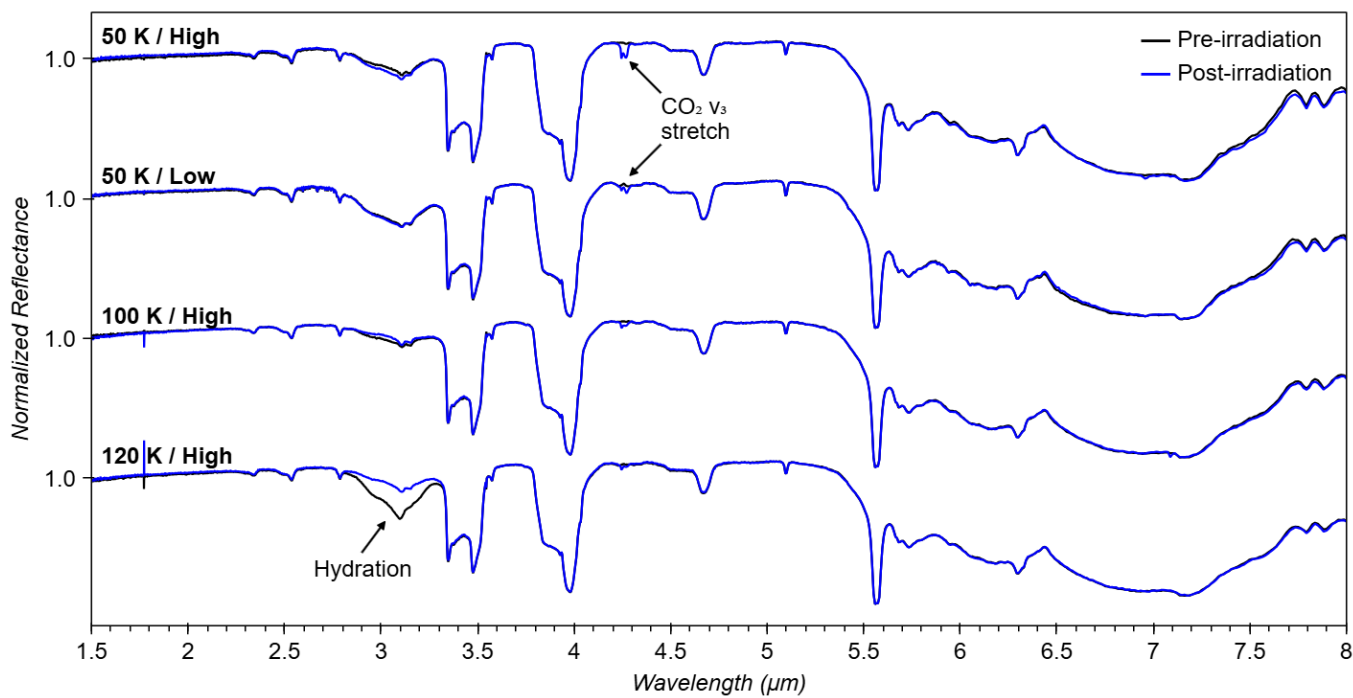

**Figure 1.** Normalized FTIR spectra of $CaCO_3$ at 50, 100, and 120 K, before and after 6 hr of irradiation with 10 keV electrons at 4244 nA cm$^{-2}$ (high flux) and 1415 nA cm$^{-2}$ (low flux). Spectra are largely unchanged upon irradiation except near 4.26 $\mu$m, where a characteristic $CO_2$ $\nu_3$ absorption doublet emerges. Weakening of the 3 $\mu$m feature indicates dehydration.

## 3. Results and Discussion

The normalized reflectance spectra of $CaCO_3$ before and after 10 keV electron irradiation for 6 hr at 50, 100, and 120 K, under both high- and low-flux conditions, are shown in Figure 1. The preirradiation spectra exhibited the characteristic carbonate absorption bands at 3.3–3.5 $\mu$m and 3.8–3.9 $\mu$m. A 3 $\mu$m hydration feature was also present, which remained constant while the sample was incubated under UHV and likely originated from intrinsic hydration of the salt sample.

Following irradiation, the spectra remained largely unchanged across most of the mid-IR region, except near 4.26 $\mu$m, where a new absorption doublet appeared. This feature corresponds to the $\nu_3$ asymmetric stretch of $CO_2$ and was most pronounced in the 50 K, high-flux spectrum. A reduction in the 3.0 $\mu$m feature was also observed, particularly at 120 K, consistent with partial dehydration of the salt during the experiment. The lack of spectral changes in the core carbonate vibrational regions upon irradiation implies minimal structural alteration of the bulk $CaCO_3$, consistent with its known resistance to radiation (I. G. Zhevtun et al. 2024). Instead, the appearance of a single new absorption feature suggests that the radiolytic chemistry induced by 10 keV electrons leads to $CO_2$ being formed and retained in the salt.

Figure 2 shows the time-resolved spectral evolution of the 4.26 $\mu$m feature associated with the $\nu_3$ asymmetric stretch of $CO_2$ for the four experimental conditions. The spectra are normalized to the preirradiation baseline and truncated to the 4.2–4.3 $\mu$m region to highlight the absorption. In all experiments, an absorption doublet consisting of a sharp component around 4.25 $\mu$m and a broader absorption near 4.27 $\mu$m appeared within minutes of irradiation, deepened with electron exposure, and eventually saturated. The relative intensities of the two absorptions varied across experimental conditions. Data in the 4.1–4.4 $\mu$m region of the normalized, continuum-removed spectra were fit with a double Gaussian profile using the SciPy, lmfit, and emcee Python packages. Band centers varied by <0.002 $\mu$m, indicating systematic error. The central wavelengths reported below are averages of these measurements.

At 50 K, the components of the doublet were well resolved, and the high-flux spectrum showed strong absorption lines at 4.246 ± 0.001 $\mu$m and 4.266 ± 0.001 $\mu$m. A low irradiation flux resulted in weaker bands. The band center of the shorter wavelength component remained the same, while the longer component shifted to 4.270 ± 0.001 $\mu$m. At 100 and 120 K, the absorptions were broader and less intense, and the two components were not as well separated. The shorter wavelength feature appeared at 4.247 ± 0.001 $\mu$m in both cases, representing a 0.001 $\mu$m shift relative to the 50 K spectra. The longer component appeared at 4.264 ± 0.002 $\mu$m at 100 K and 4.265 ± 0.002 $\mu$m at 120 K. The band center shifts suggest that $CO_2$ may be trapped in slightly different environments depending on the irradiation and thermal conditions. Overall, these observations indicate that radiolytic $CO_2$ formation and trapping begin almost immediately after irradiation starts and continue until a saturation point is reached. Both processes are favored at lower temperatures and higher electron flux.

A closer look at the band areas of the $CO_2$ absorption doublet confirmed this trend quantitatively (Figure 3). In all experiments, the total band area increased rapidly at first, then slowed down and plateaued. The highest $CO_2$ band area was observed in the 50 K/high-flux experiment (3.4 × 10$^{-3}$ $\mu$m), followed by the 50 K/low-flux (1.5 × 10$^{-3}$ $\mu$m), 100 K/high-flux (1.3 × 10$^{-3}$ $\mu$m), and 120 K/high-flux (1.0 × 10$^{-3}$ $\mu$m) experiments. In every case, the 4.27 $\mu$m component contributed more to the total band area than the 4.25 $\mu$m feature. At 50 K, the area of the 4.25 $\mu$m absorption scaled proportionally with flux, reaching $9 \times 10^{-4}$ $\mu$m at high flux and $3 \times 10^{-4}$ $\mu$m at low flux. In contrast, the 4.27 component did not show a linear correlation with flux.

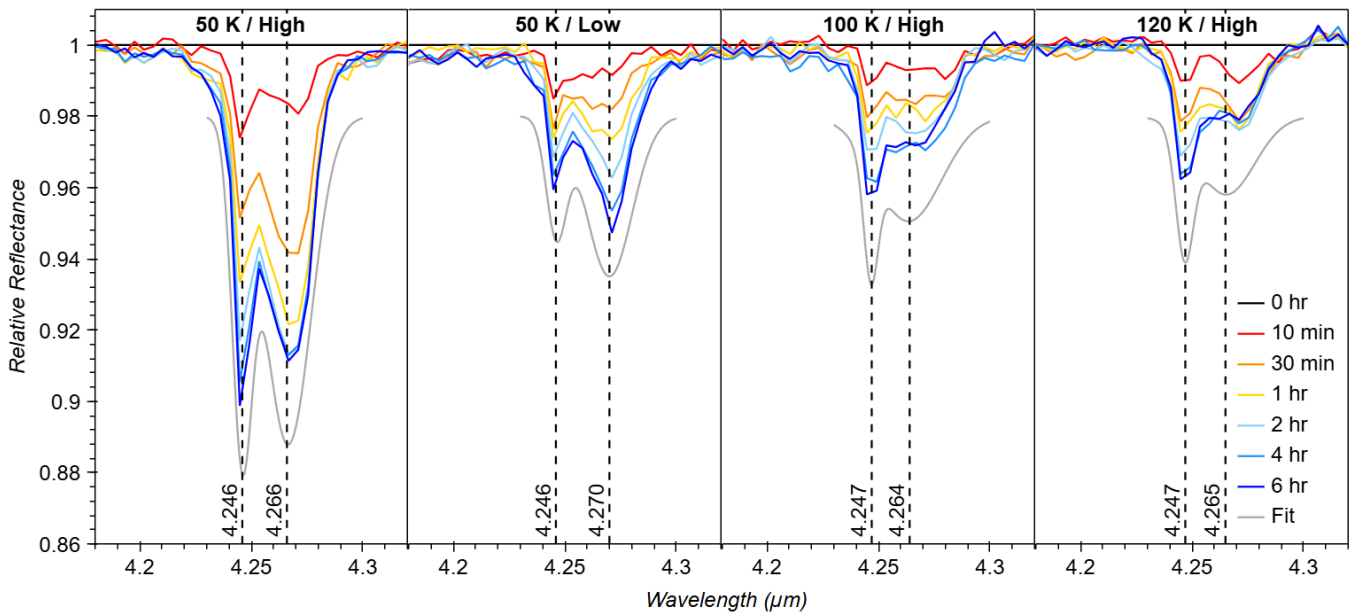

**Figure 2.** Time-resolved IR spectra from electron irradiation of $CaCO_3$ at 50, 100, and 120 K under two electron fluxes, normalized to the preirradiation $CaCO_3$ baseline and truncated to the 4.2–4.3 $\mu$m region to highlight the $CO_2$ $\nu_3$ band. Vertical dashed lines mark the central wavelengths of the $CO_2$ doublet, determined from double Gaussian fits to the continuum-removed spectra (gray). $CO_2$ absorptions appear upon irradiation and gradually saturate. The strongest absorptions occur at 50 K under high flux. At higher temperatures, the absorptions are weaker, but radiolytic $CO_2$ production is still apparent.

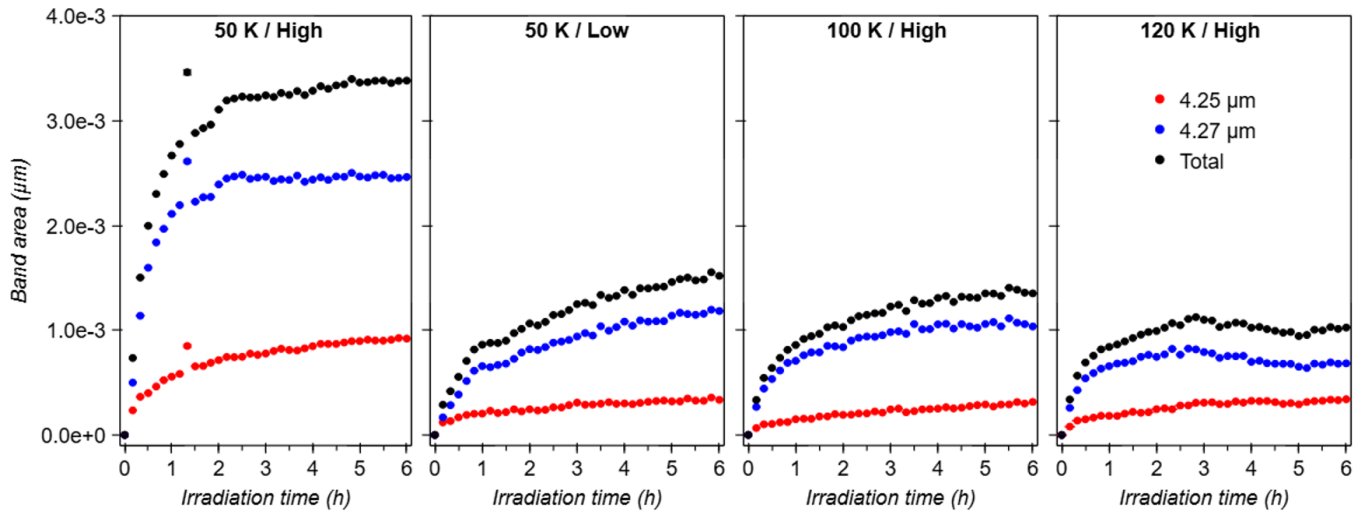

**Figure 3.** Evolution of the $CO_2$ $\nu_3$ band area while irradiating $CaCO_3$ at different temperatures and electron fluxes. Areas of the 4.25 $\mu$m (red) and 4.27 $\mu$m (blue) components, as well as their sums (black), are shown. Errors are within 0.001 $\mu$m. In all cases, the band area increases rapidly and then levels off. Total $CO_2$ yield follows the trend 50 K/high flux > 50 K/low flux > 100 K > 120 K. The 4.27 $\mu$m component is the major contributor to the band area.

At higher temperatures, the 4.25 $\mu$m band areas converged toward the same value observed in the 50 K/low-flux experiment. This means that variations in $CO_2$ band shape and intensity were primarily driven by the broader 4.27 $\mu$m feature, which is more sensitive to thermal effects. Additionally, the larger band areas observed in the 50 K experiments may reflect enhanced $CO_2$ trapping rather than production, as $CO_2$ ice is stable at this temperature and may accumulate on the surface. Thus, while it is clear that irradiation flux influences $CO_2$ production, the role of temperature in forming and trapping $CO_2$ remains unknown.

Time-resolved pressure measurements collected by the RGA throughout the experiment revealed that electron irradiation triggered an immediate rise in $CO_2$ partial pressure, which stabilized under continued exposure and returned sharply to baseline once irradiation ceased (Figure 4). The gradual decay in $CO_2$ flux is consistent with the depletion of available carbonate in the top 1 $\mu$m of the sample. The drop in the $CO_2$ partial pressure after irradiation, along with the continued growth of the $CO_2$ $\nu_3$ band, provides strong evidence that $CO_2$ is produced and trapped by the carbonate, and that the spectral signature is not from background $CO_2$ adsorbing onto irradiation-induced trapping sites.

During postirradiation warm-up, the 50 K samples released $CO_2$ near 80 K, consistent with sublimation of $CO_2$ condensed on the sample or cold finger. In the high-flux case, two additional desorption peaks appeared above 110 K, separated by roughly 20 K. The 50 K, low-flux sample produced weaker desorption signals but showed similar $CO_2$ release above 110 K. In contrast, the higher-temperature experiments showed delayed desorption, initiating around 130 K and peaking just below 150 K. A control experiment with no irradiation showed little increase in $CO_2$ with temperature. Since there was no irradiation-driven chemistry or $CaCO_3$ lattice damage in this case, the $CO_2$ would need to have been retained through other interactions, perhaps with residual water. However, the irradiated samples have much higher $CO_2$ content coming off, which makes a case for $CO_2$ being trapped in the carbonate as well. These results point to multiple $CO_2$ trapping sites with varying binding energies, present within both the carbonate and trace water.

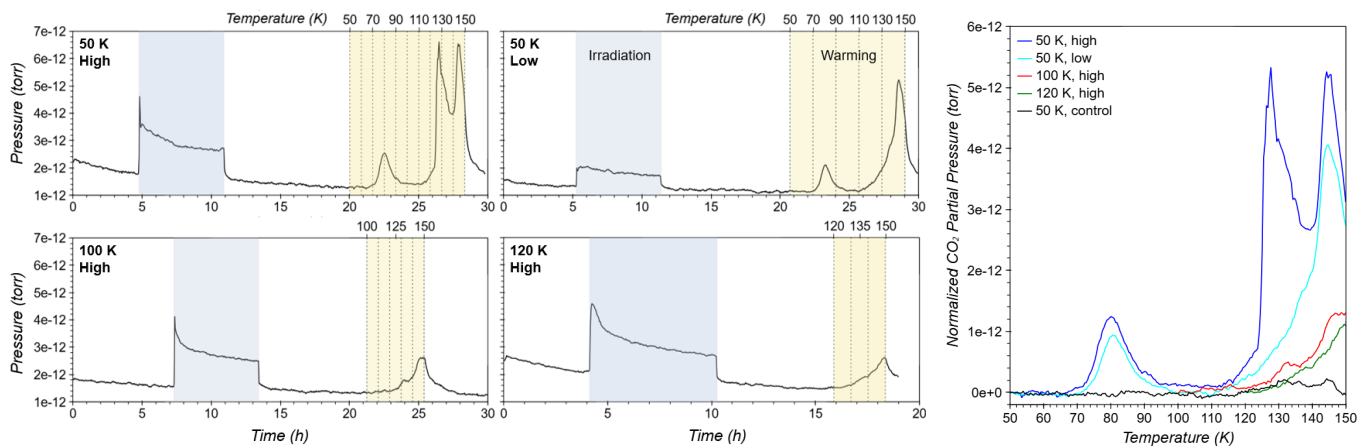

**Figure 4.** Temperature programmed $CO_2$ desorption from irradiated $CaCO_3$, monitored via RGA. Traces represent the $CO_2$ partial pressure, obtained as the sum of signals at $m/z$ 43–45. The shaded blue region denotes the irradiation period; the beige region corresponds to postirradiation warming at 0.2 K minute$^{-1}$. Irradiation induces an immediate rise in $CO_2$ partial pressure, indicating radiolysis. Upon warming, the 50 K samples show a $CO_2$ release near 80 K, consistent with sublimation of crystalline $CO_2$, absent in the high-temperature experiments. The 50 K, high-flux sample shows two desorption features above 120 K, indicating at least two trapping states. At high temperatures or under low flux, $CO_2$ release initiates near 130 K.

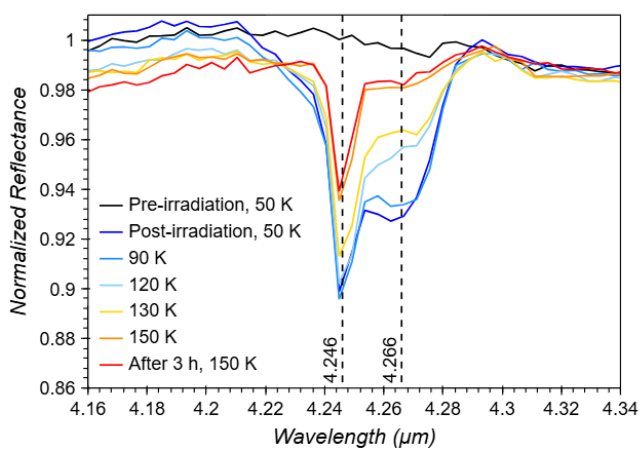

**Figure 5.** Normalized IR reflectance spectra showing the thermal evolution of the $CO_2$ $\nu_3$ doublet during warming of irradiated $CaCO_3$ from the 50 K, high-flux experiment. Dashed lines mark the fitted central wavelengths of the two components. The 4.27 μm feature weakens by 90 K and disappears by 150 K, while the sharper 4.25 μm component remains stable up to 120 K, then diminishes, but remains detectable after 3 hr at 150 K. This behavior aligns with the RGA data (Figure 4), which show a distinct $CO_2$ release peak around 90 K followed by broader desorption features above 110 K, and highlights multiple trapping states.

The RGA results for the 50 K experiments show that the irradiation flux influences both the yield and the trapping mechanisms of $CO_2$. The distinct $CO_2$ desorption peak at 130 K only appeared when the flux was high, and was accompanied by a higher $CO_2$ yield. This flux dependence may reflect kinetic competition between multiple processes such as $CO_2$ production, energy dissipation, local heating, and the creation or collapse of trapping environments. Although our experiments were not designed to isolate specific mechanisms, it is clear that flux, and not simply absorbed dose, affects $CO_2$ formation and retention.

The IR response to postirradiation thermal ramping was recorded in one trial of the 50 K, high-flux experiment (Figure 5). Spectra were collected continuously over 10 minute intervals during warming at a controlled rate of 0.2 K minute$^{-1}$, providing a temperature resolution of approximately 2 K per spectrum. The evolution of the $CO_2$ absorption doublet revealed distinct thermal behaviors for its two components: the 4.27 μm feature began to decline by 90 K, while the sharper 4.25 μm component remained largely stable up to 120 K and only weakened at 150 K, even after 3 hr at that temperature. This progression parallels key features in the RGA desorption data, which showed an early release peak around 90 K and a broader desorption phase beyond 110 K, but did not reach baseline levels even at 150 K.

Despite general similarities, the spectra and RGA results do not correspond in a simple one-to-one manner. The 90 K desorption peak corresponds to a decline in the 4.27 μm feature, whereas the higher-temperature releases above 110 K coincide with a gradual weakening of the 4.25 μm band, which persists even at 150 K. The persistence of the 4.27 μm feature beyond 80 K indicates it is not from pure crystalline $CO_2$, which sublimes at 80 K and absorbs at 4.271 μm, though its weakening near 90 K makes it difficult to eliminate partial contribution from $CO_2$ ice alongside radiolytically produced $CO_2$. On the other hand, the prolonged stability of the 4.25 μm absorption points to $CO_2$ trapped within the solid matrix. These results suggest that the spectral doublet does not represent two discrete trapping states, but rather spectrally distinguishable populations of $CO_2$ that may exist across multiple trapping environments.

The two desorption peaks in the 50 K, high-flux RGA data further support the presence of at least two such trapping environments, likely corresponding to physisorption since the $CO_2$ capture is thermally reversible. Hydration may also influence the $CO_2$ retention capacity, as seen for sodium carbonate (Y. Cai et al. 2018). Although the reported temperatures are lower limits due to imperfect thermal coupling between the grains and cold finger, any offset would imply even greater thermal stability of the trapped $CO_2$. Overall, the FTIR and RGA data provide the first evidence that low-energy electron irradiation can both generate and retain $CO_2$ in $CaCO_3$ at temperatures comparable to those on Europa's surface.

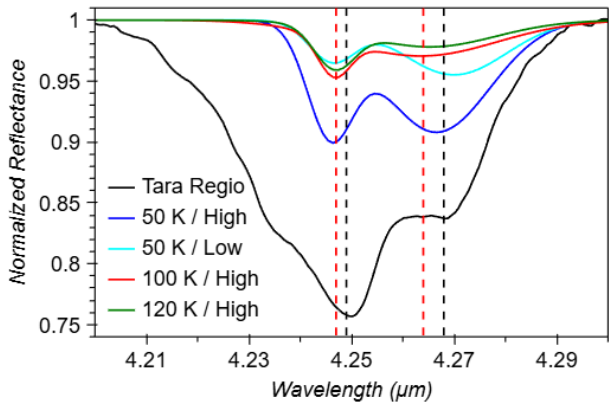

**Figure 6.** Comparison between the $CO_2$ $\nu_3$ band observed by JWST at Tara Regio and the lab spectra of 10 keV electron-irradiated cryogenic $CaCO_3$. The black line shows the continuum-normalized, five-point smoothed spectrum from the center of Tara Regio (16°S, 79°W), with dashed lines marking the band centers at 4.249 and 4.268 $\mu$m (S. K. Trumbo & M. E. Brown 2023). Overlaid are the double Gaussian fits to the postirradiation lab spectra. Red dashed lines at 4.247 and 4.264 $\mu$m mark the wavelengths at 100 K.

## 4. Conclusion and Implications

Our work constitutes the first experimental evidence of $CO_2$ generation and retention upon irradiation of calcium carbonate under cryogenic, UHV conditions, with FTIR spectra and RGA data confirming its stability at temperatures comparable to those on Europa. The development of an absorption doublet at 4.25 and 4.27 $\mu$m in the IR spectra, characteristic of the $CO_2$ asymmetric stretch, confirms radiolytic chemistry resulting in the cleavage of the carbonate group. Postirradiation thermal ramping showed that $CO_2$ was released in two stages over a broad temperature range beyond 120 K and remained spectroscopically detectable even at 150 K, implying some level of stabilization through interaction with the $CaCO_3$ matrix or minor water content. These observations suggest the presence of at least two distinct trapping environments with different sublimation temperatures, but individual retention mechanisms remain unknown.

The laboratory spectra broadly resemble the spectrum of Tara Regio in the 4.2–4.3 $\mu$m region, with overall agreement in the $CO_2$ $\nu_3$ band shape and position (Figure 6). In all cases, the absorption appears as a doublet, with individual bands at approximately 4.25 and 4.27 $\mu$m. Although the band centers vary across experimental conditions, they are close to the Europan wavelengths of 4.249 ± 0.001 $\mu$m and 4.268 ± 0.002 $\mu$m (S. K. Trumbo & M. E. Brown 2023) and overlap at the edges of the error bars. At 100 K, our measured band centers are within 2$\sigma$ of the Tara Regio values, corresponding to a blueshift of 0.002 and 0.004 $\mu$m for the

4.25 and 4.27 $\mu$m components, respectively. As for bandwidth, the 4.25 $\mu$m absorption in the lab spectra remains narrow and relatively consistent across temperatures (FWHM = 0.007–0.01 $\mu$m), comparable to that observed at Tara Regio. The 4.27 $\mu$m component, however, shows greater variability and is consistently broader than the Europan counterpart, reaching FWHM = 0.035 $\mu$m at 100 K.

Recent work by C. Goldberg et al. (2025) demonstrates that Europa's $CO_2$ band positions vary across geologic terrains, suggesting that our small wavelength offsets may not be significant indicators of compositional mismatch. These minor spectral discrepancies may stem from contrasting physical environments: our experiments use pure carbonate, while Europa's surface is predominantly water ice, which likely influences the $CO_2$ formation and trapping pathways and causes slight wavelength shifts (Y. Cai et al. 2018; L. Schiltz et al. 2024). Moreover, direct extrapolation of our laboratory results to Europa's surface chemistry is not straightforward, since Europa experiences substantially lower irradiation flux and the energy is deposited by a mix of charged particles whose distribution varies spatially and temporally. Further, 10 keV electrons penetrate only about 1 $\mu$m into the sample, whereas radiation on Europa can reach depths up to 1 m (C. Paranicas et al. 2002), processing far larger volumes of surface material over geological timescales.

Despite these caveats, the capacity of irradiated cryogenic carbonate to both produce and trap $CO_2$ is an intriguing result and expands the growing experimental library of plausible raw materials and trapping mechanisms for $CO_2$. While our findings do not allow us to determine whether all the $CO_2$ observed on Europa originates from carbonate radiolysis, they demonstrate that the process is feasible under relevant environmental conditions, and make it difficult to eliminate the possibility that carbonate salts may contribute to Europa's $CO_2$ budget. Even if carbonate radiolysis is not a dominant pathway, its inclusion broadens the scope of endogenic processes that may sustain Europa's chemically active surface.

Experimental efforts, along with high-resolution observational data of outer solar system objects, are uncovering an increasingly complex picture of $CO_2$ distribution, sources, and trapping mechanisms in icy, irradiated environments. Additional laboratory investigations are necessary to understand the trapping mechanisms observed here, test other carbonate compositions more representative of Europa's surface, and explore interactions with relevant host materials such as hydrated salts or amorphous water ice. More comprehensive spectra of the Galilean satellites, or a compilation of lab-determined optical constants for potential $CO_2$ sources, will also be critical in identifying minor components and understanding their geologic context. Together, these efforts will help clarify the origin, stability, and mobility of $CO_2$ across diverse planetary environments.

## Acknowledgments

We thank two anonymous reviewers for helping us strengthen this paper. This research was supported by grant #668346 from the Simons Foundation. A.P. is supported by the National Science Foundation Graduate Research Fellowship Program under Grant No. 2139433. Any opinions, findings, and conclusions or recommendations expressed in this material are those of the author(s) and do not necessarily reflect the views of the National Science Foundation.

Samantha Trumbo provided the reduced JWST Tara Regio spectrum associated with Proposal ID 9230. The authors gratefully acknowledge Matthew Belyakov, Ryleigh Davis, William Denman, Merritt McDowell, and Samantha Trumbo for valuable discussions about this project.

## ORCID iDs

Ashma Pandya  https://orcid.org/0000-0003-3303-1009
Swaroop Chandra  https://orcid.org/0000-0002-4960-3043
Michael E. Brown  https://orcid.org/0000-0002-8255-0545